# Lattice Percolation Approach to 3D Modeling of Tissue Aging

Vyacheslav Gorshkov,[a] Vladimir Privman,[b] and Sergiy Libert[c],*

[a]National Technical University of Ukraine — KPI, Kiev 03056, Ukraine
[b]Department of Physics, Clarkson University, Potsdam, NY 13699, USA
[c]Department of Biomedical Sciences, Cornell University, Ithaca, NY 14853, USA
*E-mail: libert@cornell.edu

**ABSTRACT**

We describe a 3D percolation-type approach to modeling of the processes of aging and certain other properties of tissues analyzed as systems consisting of interacting cells. Lattice sites are designated as regular (healthy) cells, senescent cells, or vacancies left by dead (apoptotic) cells. The system is then studied dynamically with the ongoing processes including regular cell dividing to fill vacant sites, healthy cells becoming senescent or dying, and senescent cells dying. Statistical-mechanics description can provide patterns of time dependence and snapshots of morphological system properties. The developed theoretical modeling approach is found not only to corroborate recent experimental findings that inhibition of senescence can lead to extended lifespan, but also to confirm that, unlike 2D, in 3D senescent cells can contribute to tissue's connectivity/mechanical stability. The latter effect occurs by senescent cells forming the second infinite cluster in the regime when the regular (healthy) cell's infinite cluster still exits.

# 1. INTRODUCTION

The processes responsible for aging are of great interest, but are poorly understood. Here we apply three-dimensional (3D) lattice percolation theory modeling to elucidate certain aspects of tissue aging by analyzing structures consisting of interacting cells at lattice sites. A recent study of two-dimensional (2D) models[1] identified the mechanism by which cell senescence influences lifespan, confirming experimental *in vivo* findings.[2,3] This success suggests that 2D and 3D percolation modeling incorporating cellular processes has the potential to quantify mechanisms that control longevity and tissue homeostasis (long-duration steady state). Here we demonstrate that 3D modeling allows us to explain how certain cellular dynamics properties contribute to preservation of not only cellular connectivity but also mechanical stability. We confirm the conclusion that cell senescence contributes to tissue integrity over long periods of time,[1] and additionally we report that senescence contributes to tissue's connectivity/mechanical stability, as found experimentally.[4] Interestingly, upgrading the modeling from 2D to 3D allows us to study novel phenomena; we will argue below that, unlike 2D, in 3D senescent cells can form the second, percolating infinite cluster at the same time that the healthy dividing cells' infinite cluster is also still present.

Experimentally, it was shown that several critical cellular properties correlate with longevity. For example, enhanced cellular resistance to stress: resistance to apoptosis (programmed cell death), correlates with extended longevity, both within and across species.[5,6] Cells from mammals with longer lifespans have higher rates of DNA repair,[7] resistance to transformation by viruses, and capacity to repair oxidative damage.[6] However, the extent to which these properties influence lifespan and the overall tissue "health" (integrity, connectivity, mechanical stability), and experimental verification as well as modeling of these correlations is missing. It is not known whether these or other properties contribute significantly to longevity or are merely secondary adaptations.

Processes that ensure tissue homeostasis and integrity, such as damage repair or cell replacement are a topic of a major interest to both medical professionals[8] and basic scientists.[9] In order to identify the parameter relations that ensure tissue homeostasis for extended periods of time, in our lattice-connectivity percolation modeling approach we incorporate time dependence mimicking known cellular processes: cell division, senescence, apoptosis (programmed cell death), etc., and numerically obtain cellular statistics and cluster-connectivity evolution with time. Generally, percolation models can provide information and predictions on the system's integrity and general connectivity.[10-12] Our modeling is to an extent reminiscent of earlier percolation approaches to autonomous self-healing and self-damaging in "smart" materials,[13-16] inspired by biological properties of tissues.

Tissues are studied dynamically, incorporating a selection of ongoing cellular processes, depending on the complexity of the model. These can include apoptosis, cell division to fill "vacant sites" left by dead cells, cell senescence, senescent cells disrupting functioning of adjacent healthy cells, etc. The connectivity information obtained for different cell types can then in principle be related to physical or biological tissue properties and structure. In addition to aging, such models can also offer insights into processes that affect local damage (wound) repair or recovery after systemic damage (e.g., chemical insult). Cells are placed on a lattice and Monte Carlo (MC) sweeps alter the system according to a probabilistic set of rules. Briefly (details will be given later), such rules can include the following dynamics: Regular cells can divide, but the number of cell divisions cannot exceed the Hayflick limit,[17,18] upon reaching which cells undergo senescence. Senescent cells permanently withdraw from cell cycling (i.e., they do not divide). Note that senescence can be triggered[19] by either exhaustion of a cell's division potential or by damage. Cell damage could be extrinsic (physical or chemical insult) or intrinsic (inflammation or DNA mutation). Severely damaged cells die; moderately damaged cells can either enter growth arrest (temporary senescence), permanent senescence, or apoptosis.[20] Apoptosis can be triggered by irreparable damage or severe insult.[21] Healthy (dividing) cells have a certain rate of apoptosis; they might be sensitized to apoptosis by inflammation. Senescent cells are resistant to apoptosis,[22] "dying" in MC sweeps at a significantly reduced rate. Here we only consider a subset of all such possible dynamical rules.

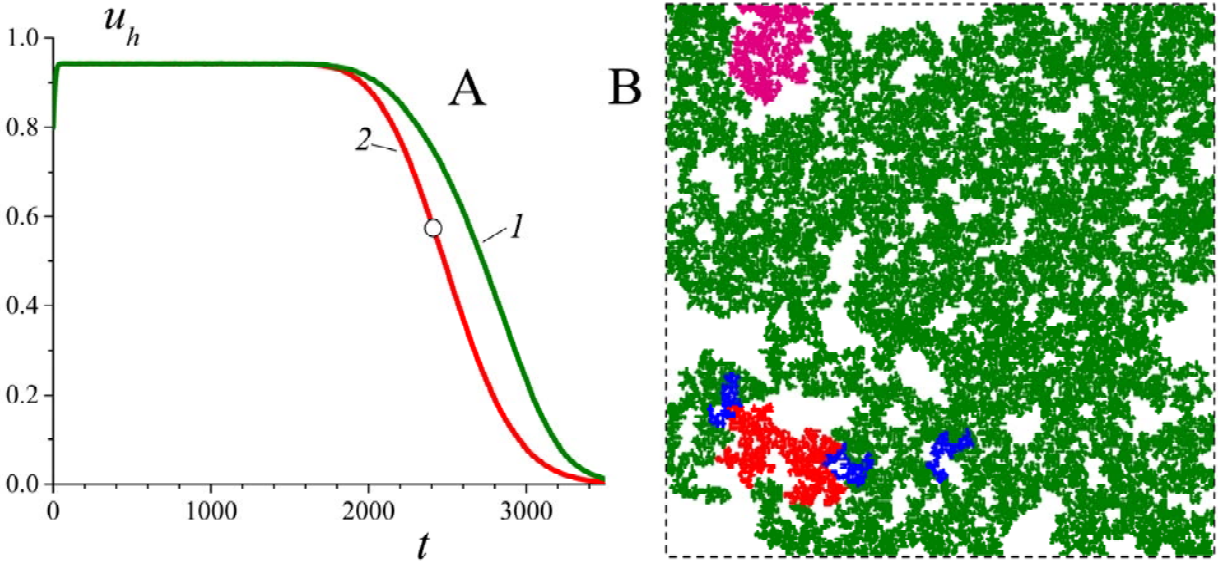

**Figure 1**. Illustrative results for a single-layer 2D square-lattice model, similar to Ref. 1, but here taken with the same details of the dynamical rules as in our 3D modeling, defined Sec. 2 and 3. Here the initial healthy cell density (fraction) was $u_h(t=0) = 0.8$, cell death rate $p = 0.01$, cell division rate $q = 0.05$, cell division potential was exhausted after 50 divisions and then they became senescent. (A) Line *1* (green): tissue integrity measured as the density of healthy cells, $u_h(t)$, when senescent cells die at the same rate as healthy cells. Line *2* (red): The same when senescent cells are present at greater density because they were set to resist apoptosis (here, made immortal). Note the similarity of these curves to the classical Kaplan–Meier survivorship plots in aging research.[23-26] (B) A large-cluster structure snapshot obtained in a random MC run at a point marked by the open circle in panel A. Color-coding here represents the three largest (in the count of cells/sites) *connected* healthy-cell clusters, color-coded: green > red > magenta, and the sites in the three largest connected senescent-cell clusters: all in blue. The point (in panel A) for which the cluster structure is shown in B, was selected at the time just before the largest cluster of healthy cells (green) lost its percolation property of spanning the system in both directions; here $u_h = 0.57$, whereas the percolation connectivity is lost at a somewhat later time for which $u_h = u_h^{cr} \simeq 0.55$ for this selection of the model parameters.

Results of one of our 2D modeling studies are illustrated in Fig. 1. With proper choice of parameters this system reproducibly establishes a long-duration steady-state (homeostasis) followed by the connectivity loss (aging) at later times. Generally, we find that tissue integrity, dynamically assessed as the time-dependent fraction of sites with healthy cells, $u_h(t)$, in percolation modeling, behaves qualitatively similarly to classical survivorship plots.[23-26] Indeed, it is expected that survivorship of the organism should be proportional to the health and integrity of its tissues, and probability of death should increase as individual tissues deteriorate and lose functionality. Interestingly, when we added the rule that senescent cells resist apoptosis (by making them immortal, effectively increased the number of senescent cells in the system) we obtained a curve indicating reduced longevity: compare *1* to *2* in Fig. 1A. This observation qualitatively supports recent experimental findings that inhibition of senescence,[2,3] e.g., by linking[2] a senescence marker gene to the activation of apoptosis, results in extended longevity and amelioration of age-associated pathologies.

Earlier studies of percolation models that involve various types of "cells" on a lattice, which interact with each other resulting in the change in their cluster structure and therefore connectivity and various physical properties as functions of time, *t*, were carried out in response to a recently increasing interest in "smart materials", which utilize nanoscale features to achieve useful properties, such as autonomous self-healing.[13-15,27-36] In such materials, development of damage and fatigue can be delayed by embedded capsules containing a microcrack-healing agent activated by a local "triggering" mechanism.[15] Very recently, there has also been interest in materials with autonomous self-damaging properties,[16,37,38] also termed self-destructive, self-deteriorating, or transient (bioresorbable), as well as in situations when both mechanisms are utilized.[16] The first autonomous self-healing polymer composite was realized[31] with the polymerization process initiated by the released healing agent preventing the propagation of cracks caused by mechanical stress. This finding was followed by many interesting experimental and theoretical developments. Experimental realization of "transient (bioresorbable) electronics" at the device-component level aimed at injectable devices has been reported very recently.[37,38]

The aforementioned advances in materials and device sciences are obviously *bio-inspired*. Therefore, it is natural to also apply lattice percolation modeling ideas to actual biological tissues and cell cultures. While the modeling is similar to approaches used for materials,[13-16] our preliminary 2D studies[1] indicate that there are obvious and significant differences. Indeed, in tissues each cell is an active element, changing with time and also affecting other cells. The modeling approach is described in Sec. 2. Section 3 addresses new features found in 3D as compared to 2D, whereas Sec. 4 offers comments to visualization of 3D clusters, which is of course less straightforward than the 2D illustration in Fig. 1B. Long-duration steady state formation is addressed in Sec. 5. Section 6 describes percolation properties of senescent cells. Then, in Sec. 7 we comment on the nature of the various percolation transitions in the considered system. Section 8 offers a summarizing statement.

## 2. THE MODELING APPROACH

We idealize a part of a tissue as 3D or 2D lattice, the latter, for instance, for layers such as in skin. Since we are interested in behavior at relatively high occupancy, size and simulation-box-boundary effects are expected to be mostly negligible for large enough system.[39] However, we also report results for percolation transitions of interest, see Sec. 7, for which care must be exercised. Various cell (lattice site) populations in the simplest modeling approach include regular healthy cells, senescent cells, and vacant sites. Each MC sweep through the system involves sites being probed at random, at the rate such that on average all the lattice sites are visited once during each unit time step. This defines MC time step units. Averages over several runs are done for estimating the cluster-structure and connectivity statistics.

Detailed substantiation for possible rather complicated transition rules and references to experiments that suggest these rule have been presented in Ref. 1, by using a 2D example of a cell layer in skin: the basal layer of epidermis (stratum basale), which is a monolayer of keratynocytes. We will not review all these arguments here. For our purposes it suffices to focus on the following simple set of rules. Initially, a finite cubic lattice, for instance, of equal sizes $N \times N \times N$ in all three dimensions, is randomly populated with healthy cells at density (fraction of the total count of lattice sites) $u_h(0)$, while all the other sites are vacant. Healthy cells can die with probability $p$ when probed in a MC sweep. They can also divide with probability $q$ to create a daughter cell that randomly fills a nearest-neighbor vacant site. However, if there are no nearby vacancies then the division is suppressed: $q$ is replaced by 0 (crowding by neighboring cells is known to suppress cell division[40]). Otherwise (with probability $1 - p - q$) the cell is unchanged.

Each healthy cell has an index of possible divisions stored for it, which is decreased by 1 after each division (and the daughter cell gets the same reduced index as the parent cell). Here the initial value of the possible divisions was taken as 50, which is a typical value of the Hayflick limit.[17,18] The nature of this limit lies in the inability of the cell to fully replicate the ends of the chromosomes (telomeres), which eventually shorten to the limit at which cell senescence is triggered.[19,41] Once a cell reaches the count 0 of possible divisions, it thus becomes senescent. Besides not dividing, these cells have several other known properties as well as effects on the behavior of other cells that have been considered to be important in tissue and organism longevity, healing ability, resistance to cancer, etc.[9] For our purposes it suffices to focus on the property that senescent cells are resistant to apoptosis,[42] i.e., they persist longer than regular cells. To consider this effect we report modeling with the probability, $p_s$, for senescent cells to die taken as $p_s = 0$, $p/3$, or $p$ (here $p$ is the healthy-cell death probability defined earlier).

Certain modeling results,[1] obtained with the homogeneous-system modeling rules for 2D, with the minimal set of dynamical rules just listed, and shown in Fig. 1, also apply in 3D, as illustrated later. For initial conditions with sufficiently large density of healthy cells the model can reproduce a long interval of stable mostly healthy cell population: the homeostasis, manifested as steady-state dynamics, i.e., persistence of the region of the flat healthy-cell density. Furthermore, as already mentioned we qualitatively confirm recent experimental findings[2,3] that inhibition of senescence, i.e., making senescent cells die at the same rate as healthy cells can extend longevity. However, new interesting features are found in 3D, as described in the following section.

## 3. NEW FEATURES SPECIFIC TO 3D

Illustrative results showing interesting new features of the considered model in 3D are shown in Fig. 2. We denote the time-dependent density of the healthy cells as $u_h(t)$, and senescent cells as $u_s(t)$. Here we took the heathy-cells death rate $p = 0.01$, and their division rate (when not blocked by neighbors) $q = 0.05$, with the count of possible divisions 50, after which senescence sets in. The latter cells die at rate $p_s = p/3$. Initially all lattice sites were filled with healthy cells, i.e., $u_h(0) = 1$, but steady state of a somewhat reduced density was rapidly established. Once division potential of a large enough fraction of the healthy cells is exhausted, the density of these cells decays and senescent cells begin to form, with their density peaking for a certain interval of times before these cells also die out. All these features are seen in Fig. 2A. Interestingly, in 3D, unlike the single-layer 2D models, the latter density, $u_s$, can reach large enough values while the density of the healthy cells, $u_h$, is also still large enough to have *two infinite percolation clusters* coexisting.

Connected clusters of same-type: healthy, senescent cells (and we also looked at vacant lattice sites) were identified as part of our simulation by using the standard algorithm.[43,44] Generally, in 3D, unlike 2D, more than a single infinite cluster can exist. This property in 3D has been explored mathematically, but only in few cases was it of relevance for the actual applications of the percolation/cluster-connectivity models.[45] The density of those sites that are in the largest connected cluster of healthy cells, $\rho_h$, drops to zero at some finite time (this is a percolation transition), past which the infinite connected cluster of healthy cells no longer exists. This is shown in Fig. 2B. Interestingly, in 3D the infinite cluster of senescent cells can form before that time, with its density (as a fraction of the total number of sites) increasing from zero at the first percolation transition of these cells, peaking, and then dropping back to zero at a later time, at the second percolation transition for senescent cells.

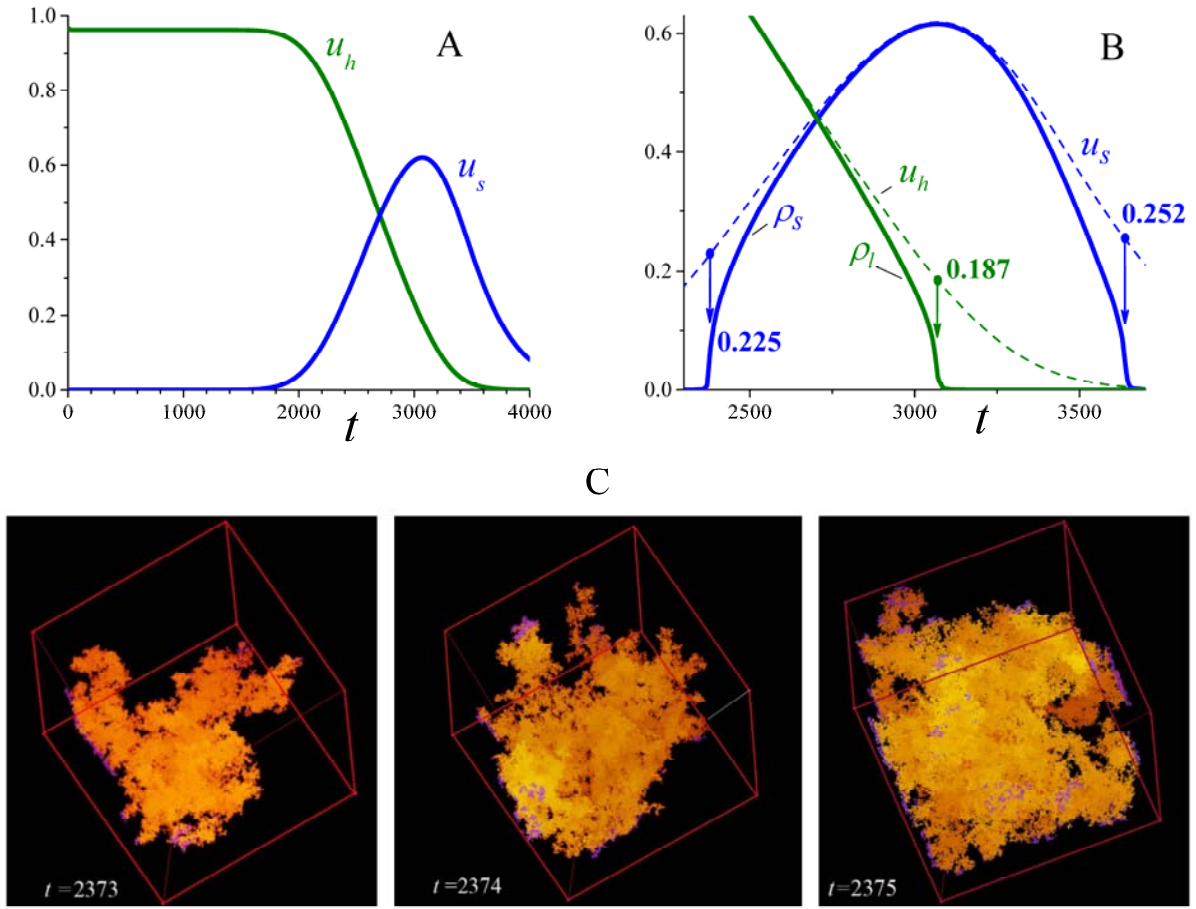

**Figure 2**. (**A**) The densities of the healthy cells, $u_h$, and senescent cells, $u_s$, for the 3D model with parameter values $p = 0.01$, $q = 0.05$, $p_s = p/3$. Other details are given in the text. (**B**) The dashed lines show the same data as in panel A for the evolution at times somewhat past the steady state regime. The solid lines show the densities of those sites that belong to the largest connected cluster of healthy cells, $\rho_h$, and the largest connected cluster of senescent cells, $\rho_s$. The total site densities, $u_{h,s1,s2}^{cr}$, at which the percolation transitions occur are marked by arrows. (**C**) Snapshots of the largest *connected* cluster of senescent cells for a random MC run, for three consecutive time steps during which it becomes "percolating," connecting all the six faces of the simulation region.

Thus, in 3D, even before healthy cells lose their infinite connectivity, senescent cells can for some time reach densities large enough to establish their own infinite cluster that spans the system. Therefore, they can contribute to the mechanical integrity (and transport properties: e.g., thermal and electrical conductivity) of the tissue, as experimentally found.[4] For a certain time interval interpenetrating infinite clusters of healthy and senescent cells coexist. The total site densities, $u_{h,s1,s2}^{cr}$, at which these transitions occur are not equal (marked by arrows in Fig. 2B) and are noticeably lower than for the uncorrelated-site 3D percolation, because our model has added local correlations (daughter sites are neighbors in contact with the parents and share the division count, means, likely retain the contact also if both survive until reaching replicative senescence).

For considerations of "tissue integrity" and various properties based on large-scale connectivity of tissue types we are interested in regimes in which infinite clusters and overall cell densities are not too small. As a result, the critical behavior at the percolations transitions *per se* is of a secondary interest in studies of the type reported here. We note that data such as those shown in Fig. 2A and B represent averages over typically 15 runs, which allowed us to use large system sizes, $N^3 = 300^3$ (with free boundary conditions). This suffices to make finite-size and boundary effects, as well as statistical noise negligible, except perhaps very close to the transition points at which $\rho_h$ or $\rho_s$ touch zero. For the study of the percolation-transition properties (Sec. 7) the number of independent runs was increased to at least 200. While careful study of the percolation critical behavior is outside the scope of the present work, we report evidence (Sec. 7) that the *universality class* of the transitions is not changed (because of the short-range nature of the added correlations). Specifically, the standard 3D critical exponent $\beta \approx 0.42$ was found for the various percolation transitions considered.

## 4. IMAGING CLUSTER MORPHOLOGY

We just noted that for tissue "health" and substantial connectivity we are interested in model regimes of dense cell-type populations, away from the percolation transitions. Therefore, one could argue that averages obtained within the mead-filed rate equation approach should suffice as good approximations. Indeed, derivation of such rate equations have been considered in the context of similar models for self-healing materials,[13,15] mentioned earlier. Lattice modeling, however, provides an added insight into the cell-cell "exclusion" effects affecting cell-type clustering and other morphological properties. Figure 2C illustrates snapshots of cluster morphology which reveal interesting insights on the system dynamics. Here the images show the rapid onset of added connectivity mediated by senescent cells as their largest cluster spans the system (an infinite cluster emerges) during those MC time steps that take the system through the first of its two senescent cells' percolation transitions. Imaging cluster morphologies in 3D is an interesting and challenging research field on its own.[46-50] An illustrative *movie* showing spatial views of the just-formed infinite cluster (the one snapshotted in Fig. 2C) is available as Supplementary Information: click the arrow in the box below to play the movie.

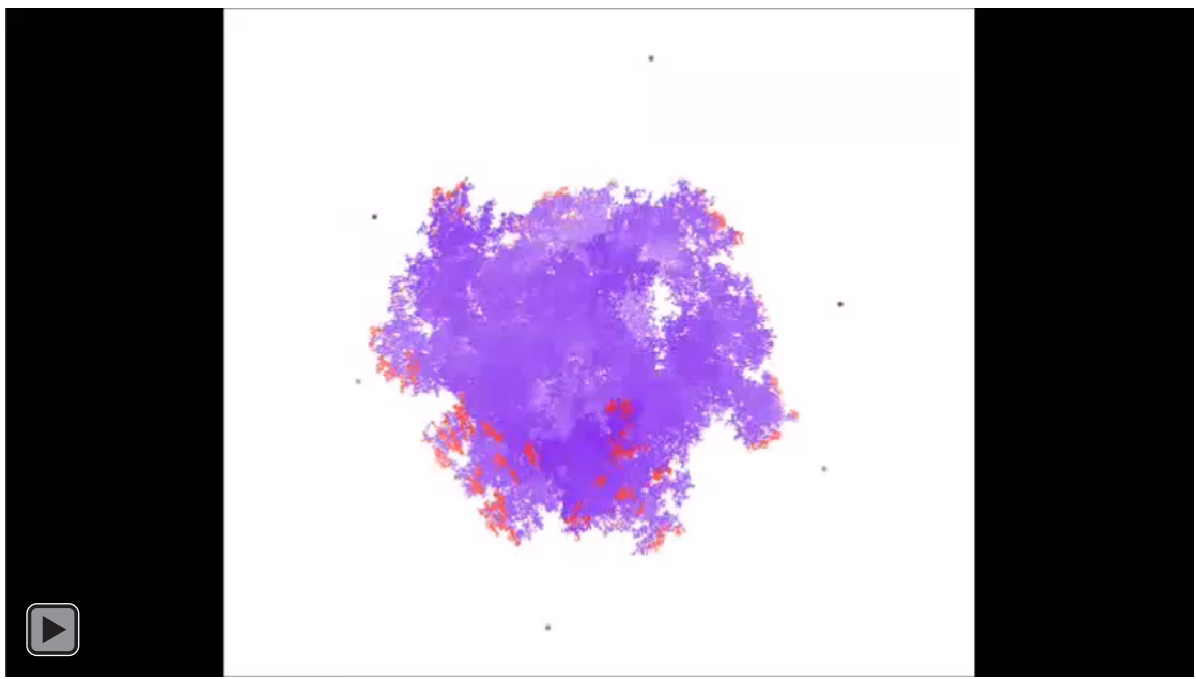

While lattice models considered here are "cartoon" in their description of the actual cell morphology and cell-cell interactions, they can capture local correlation effects, such as cell clustering and long-range connectivity. This is generally true for a description provided by lattice statistical mechanics modeling. Successful statistical-mechanics approaches to aging processes *within* cells and their relation to cancer—avoidance of which is generally intimately connected with processes that also control aging[51]—have been recently reported.[52,53] On the cellular scale, our proposed lattice-percolation approach is a complicated version of cellular automata, which is a field of study of pattern emergence. Certain other applications of percolation modeling to biological problems are reviewed in Refs. 12, 54. For many of such studies visualization of the cluster connectivity is as important as evaluating average quantities.

## 5. LONG-DURATION STEADY STATE AND THE ROLE OF VACANCIES

Establishment of the long-duration steady state (homeostasis) is a key feature of the considered dynamics at short to intermediate times. Approximate considerations can be offered to estimate the steady-state value $u_h(t) \approx u_h^{st}$. As long as we have $p < q$ (the death rate is lower that the division rate), the value of $u_h^{st}$ does not much depend on the initial state when the homeostasis develops and then persists for some time until the cell division potential is exhausted, which happens if we initially have a large enough healthy-cell density. This is illustrated in Fig. 3. Indeed, in steady state, well before the onset of senescence, the balance between cell death and their replenishment by division can be approximately represented as

$$pu_h^{st} \simeq p_v(1 - u_h^{st}), \qquad u_h^{st} \simeq p_v/(p + p_v), \tag{1}$$

where $1 - u_h^{st}$ estimates the fraction of the vacant sites, whereas $p_v$ is the average rate at which vacancies are filled by daughter cells during cell division. Denoting by $z$ the lattice coordination number ($z = 6$ for the simple cubic lattice), and assuming that the steady state is rather dense, we expect that most vacancies are surrounded by $z$ (or not much less than $z$) healthy cells and therefore

$$p_v \simeq [1 - (1-q)^z]. \tag{2}$$

For example, for $p = 0.01$ and $q = 0.05$, these relations suggest $u_h^{st} \simeq 0.963$, whereas numerically, cf. Fig. 3, we get $u_h^{st} = 0.962$.

If the aforementioned conditions are not applicable, specifically, if the initial life-cell density and/or the division rate $q$ are small enough, such that steady state, if at all formed, does not have mostly isolated vacant sites, then Equations (1)-(2) will not be applicable. For the value of $q = 0.0125$ in the above example: see Figs. 3 and 4, a persistent (for some time) steady state is formed, but the vacancies form large clusters (the average number of healthy cells around each vacancy is notably less than $z = 6$). In fact, for this value of $q$ the vacancies are close to their percolation transition of forming an infinite cluster. If we denote the fraction of the sites that are vacant as $u_v = 1 - u_h - u_s$, then our estimate at the percolation transition (formation of the infinite cluster) for this quantity is $u_v = 0.272$ (considering steady-state vacancy-cluster statistics for varying $q$, with all the other parameters fixed). The values of the percolation thresholds and the percolation transition criticality are discussed in the next two sections.

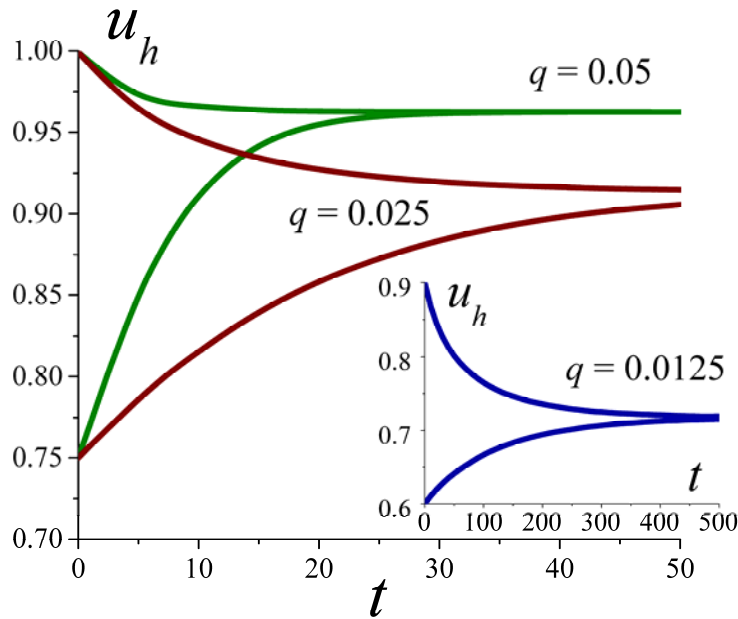

**Figure 3.** Short-time behavior showing the onset of the steady state for three choices of the value of $q$, all exceeding the values of $p$, here $p = 0.01$, each plotted for two different initial values of $u_h(0)$. The value of $p_s$ is immaterial in this case because for the shown time scales there are practically no senescent cells.

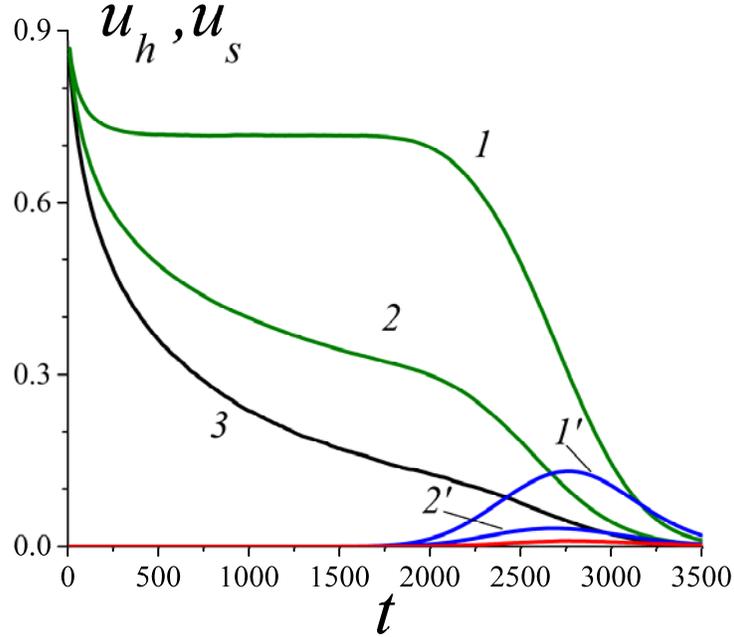

**Figure 4.** Example of a system with $p = 0.01$, $p_s = 0.01$, and $u_h(0) = 0.9$. Here the time-dependence is shown for $q = 0.0125$ (curves *1,1'*) and $q = 0.01$ (curves *2,2'*). The monotonic curves (green): healthy cells, the peaked curves (blue): senescent cells. For comparison, a 2D result for $q = 0.01$ is also shown: curve *3* (black) for healthy cells, and the lowest peaked curve (not numbered, color: red) for senescent cells.

We mentioned earlier that, at least as far as the density of the healthy cells goes, we are interested in large enough values ("viable" tissue) such that the percolation transition is not important to focus on for understanding the physical properties of the system. However, interestingly, we found that the percolation transition of vacancies in the context of the steady-state persistence may be of relevance. Figure 4 illustrates this for $q = 0.01$, somewhat lower than the approximate borderline value 0.0125. Here no persistent steady state is found at all. While it is not clear whether this can be considered as indicative of a precise correlation with the formation of the infinite cluster of vacancies, our observations are suggestive of that, once such a cluster is established, it impedes and likely prevents steady state formation/persistence even though healthy cells are still well-connected (and for some time remain above their own percolation transition density of loss of system-wide connectivity). This is also the regime where the differences between the 3D and single-layer 2D systems are profound (see Fig. 4), because of the lower connectivity of all cells types in 2D and impossibility of having interpenetrating infinite clusters in single-layer 2D systems.

## 6. THE ROLE OF SENESCENT CELLS

Generally, here we are interested in "viable tissue" situations with long-duration steady states that persist as long as cell division potential is not exhausted. As more and more cells become senescent, the fraction of the healthy cells will decay to zero. Figure 5 shows a 3D simulation with a choice of parameters that gives a long-duration steady state. Here $q = 0.05$, $p = 0.01$, and we took the initially fully occupied lattice: $u_h(0) = 1$. The system rapidly reaches steady state, and the decay of $u_h(t)$ only sets in for large times. The details of the decay depend on the senescent-cell death rate $p_s$.

As in 2D, we find that in 3D senescent cells affect the "viability" of the healthy (regular) cells, and interestingly, longer-surviving senescence (smaller $p_s$) results in shorter "life span," i.e., faster decay of $u_h(t)$. This property was discussed in detail in the 2D study,[1] and matches experimental findings.[2,3]

However, as already pointed out earlier and illustrated in Fig. 5, the density of senescent cells in 3D (unlike in 2D) can reach large enough values to form an infinite cluster well before the infinite cluster of healthy cells disappears. In fact, for the system shown in Fig. 5 the 3D percolating cluster of senescent cells sets in when the density of the healthy cells is still rather large, $u_h(t) \approx 0.72$, irrespective of the value of $p_s$. This important property corresponds to experimental observations[4] and indicates that in 3D, unlike 2D, senescent cells can contribute to system-wide mechanical and transport properties.

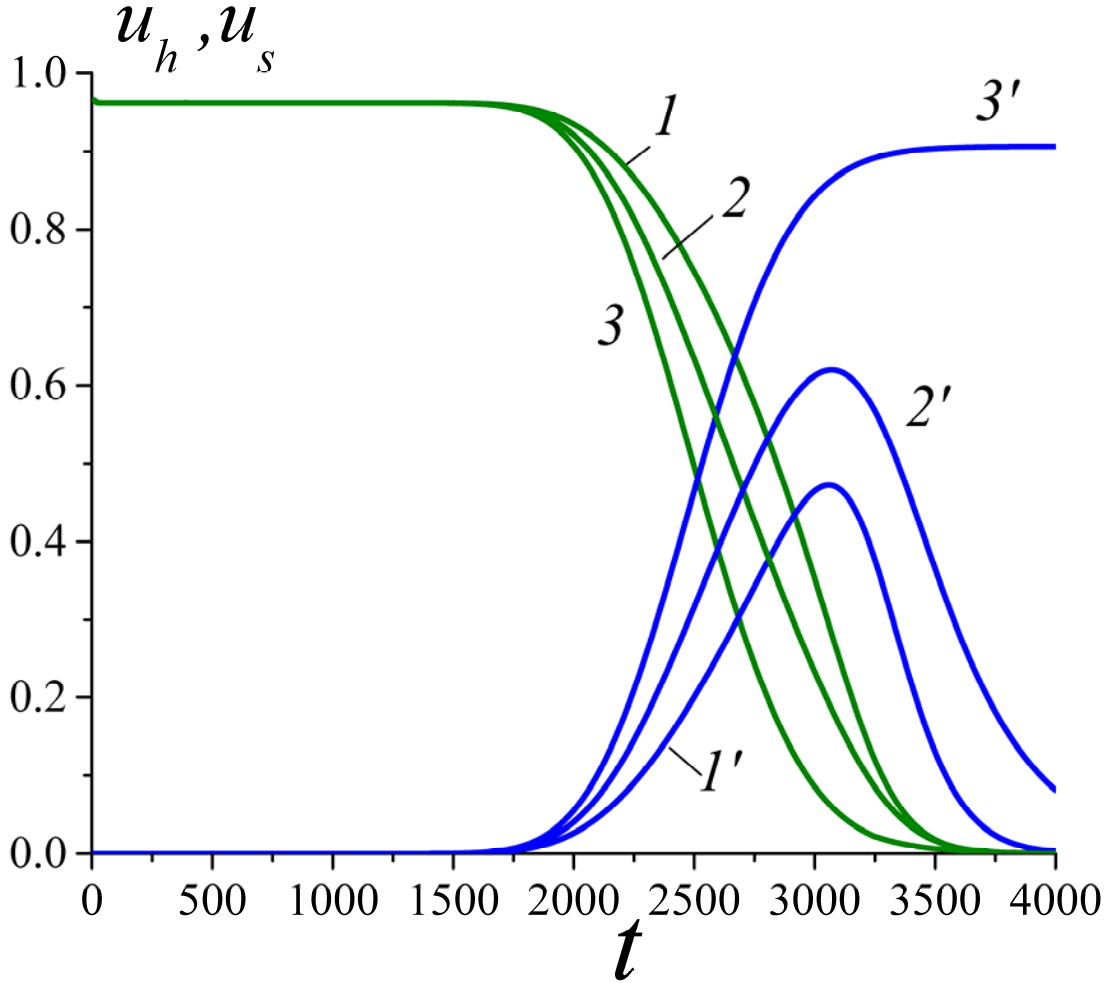

**Figure 5.** A 3D system with $p = 0.01$, $q = 0.05$, and $u_h(0) = 1$. Here the time-dependence is shown for three different values $p_s = p$; $p/3$; 0: curves *1,1'; 2,2'; 3,3'*, respectively. The monotonically decreasing curves (green) show $u_h(t)$, whereas the curves emerging for large time (blue) show $u_s(t)$. Note that the case $p_s = p/3$ was shown in Fig. 2.

We note that various percolation thresholds (reported in the preceding section and in Fig. 2) here are noticeably lower than the uncorrelated 3D site percolation threshold, which is estimated[55] as approximately 0.3116. This is obviously related to the fact that there are strong local correlations: Healthy cells cluster because daughter cells born by division are placed nearby. Since they also inherit the count of the remaining divisions, this leads to a certain degree of "inherited" local correlation in the formation of senescent cells, especially at the time of the formation of their infinite cluster, and to some extent preserved when this cluster disappears at a later time. The local correlation of vacancies is the least direct, resulting from that clusters of vacancies are only eroded at their boundaries when neighboring healthy cells divide.

Therefore, for model parameters that yields otherwise similar densities, heuristically we expect that the percolation thresholds are ordered as follows (in a self-explanatory notation):

$$u_h^{cr} < u_{s1}^{cr} \text{ (appearance)} < u_{s2}^{cr} \text{ (disappearance)} < u_v^{cr} < \text{uncorrelated sites.} \qquad (3)$$

We note that if we start with low enough healthy cell density, but with good division potential, then healthy cells can undergo another percolation transition of forming an infinite cluster at short times. Similarly, vacancies can have mode then a single transition, etc. The shown relations in Equation (3) pertain to transitions of interest as per our discussion earlier in this section.

Generally, perhaps not mathematically rigorously, but heuristically we can expect that if the lattice is sufficiently "connected" to have certain cell types percolate at densities (fractions of sites) less than 1/2, then two interpenetrating infinite clusters can coexist, and even more than two, if percolation thresholds are lower. Such expectations are obvious for uncorrelated site-occupancy, but our model has local correlations, and therefore careful investigation is needed for each particular combination of parameters.

Earlier, we emphasized that healthy- and senescent-cell infinite clusters can coexist in 3D, but not for the single-layer 2D lattice. However, for the 2D "slab" geometry $(N \times N) \times M$, with very large $N$, but with small $M = 1, 2, 3, ...$, it is expected[56] that the percolation threshold drops below 1/2 for $M > 1$ in single digits. This of course depends on the details of the model, and in our case, considering the local correlations that result in generally low thresholds, we estimate that typically for $M \geq 2$ the thresholds will all be below 1/2. However, this does not *per se* guarantee that an infinite dense cluster of senescent cells can coexist with an infinite cluster of healthy cells for some extended interval of times. We did not investigate this matter in detail, but we found preliminary evidence that rather thin, few-layer "slabs" of tissue are de-facto three-dimensional as far as their healthy- and senescent-cell connected-cluster structure goes.

## 7. THE PERCOLATION TRANSITIONS IN 3D

While the study of the nature of the percolation transitions in the present model is not central to our main investigation, we did carry out checks, described in this section, that the universality class of the considered transitions is that of the standard uncorrelated 3D percolation. This conclusion is in itself not surprising, because the correlations introduced by the model dynamics are all short-range, as described earlier. Therefore, no systematic effort at considering finite-size and surface effects was made. Instead, we verified the scaling relation

$$\rho \approx R|u - u^{cr}|^\beta , \qquad (4)$$

by a simplified approach sketched below, confirming that the critical exponent $\beta$ has the standard percolation value. Here the fraction of sites of a particular type, $u$, can be $u_h$, $u_s$ (or $u_v$ for vacancies), with their respective percolation transition values $u^{cr}$ at one or more transition points. The density of those sites that are in the infinite cluster is non-zero only on one side of the transition (in the infinite-system limit), and vanishes according to Equation (4).

Our aim was to check that the values of $\beta$ at all the transitions are consistent with the uncorrelated percolation value, estimated[57] in the literature as approximately $\beta = 0.418$. Note that the quantities $\rho(t)$ and $u(t)$ are actually functions of time, which we used to evaluate $\rho(u)$, though in principle we could also try to fit $\beta$ from $\rho \propto |t - t^{cr}|^\beta$ instead. We preferred the former approach because the values of $u^{cr}$ are of interest on their own, as discussed earlier. For these calculations we used system size $N^3 = 300^3$ as before, with free boundary conditions, but for the selected parameter combinations for which estimation of $\beta$ was attempted, we ran large-scale simulations to have averages over at least 200 runs, to minimize statistical noise.

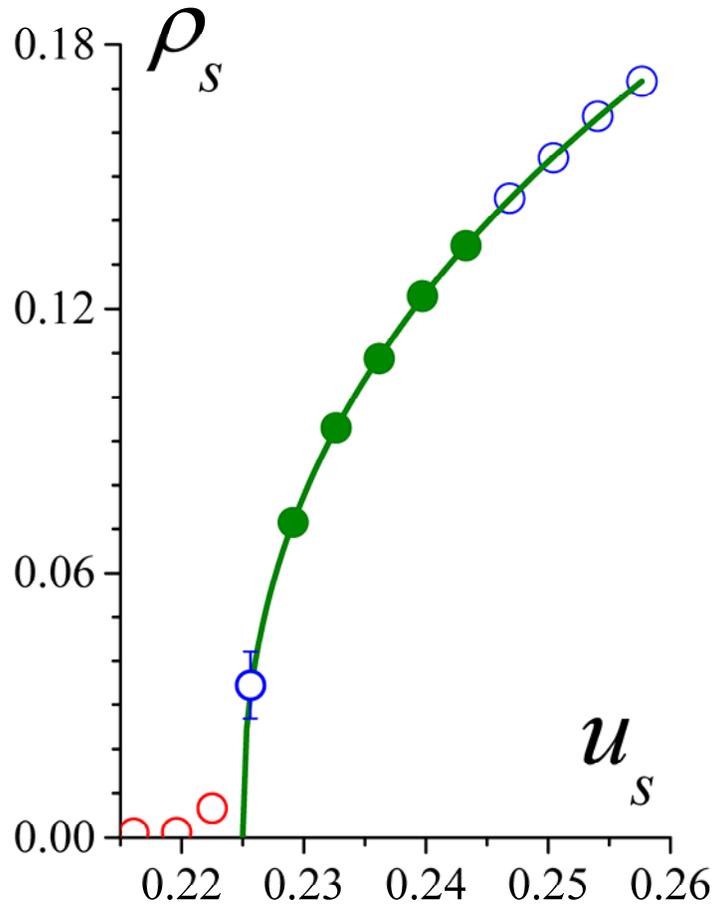

**Figure 6.** Illustration of the approach used to estimate the percolation transition parameters, here for the transition of the onset of percolation (the first transition) for senescent cells in the case of the model parameters used in Fig. 2 and 5. Full circles (green) mark the region of values used to fit (solid green curve) the parameters in the relation $\rho = R|u - u^{cr}|^\beta$, yielding estimates $R = 0.730$, $u^{cr}_{s1} = 0.225$, $\beta = 0.423$. The actual statistical-noise spread of the data (200 runs) close to the transition is shown by the error-bar at the lowest open-circle (blue) data point that is very close to the transition on the percolating site of it. The statistical fluctuations were actually found to be the largest within approximately this distance near the transition point.

The approach consisted of least-squares data fits to identify a region on the percolating-side of the transition not too far from the transition point, but also not past it, that yielded the most consistent fit to a simple power low. This is illustrated by color-coded points in Figure 6. The details of this analysis are not reproduced here, because it is not a systematic approach to eliminate finite-size and surface corrections. However, considering the large system size used, we estimated that the obtained values are accurate within about 1.5%. This was verified by applying the same procedure to the uncorrelated simple-cubic-lattice site percolation, yielding the results $u^{cr} \approx 0.313$, $\beta \approx 0.423$. The best literature estimates cited earlier were $u^{cr} \approx 0.3114$, $\beta \approx 0.418$. We note that if we actually use the literature value of $\beta$, then the same procedure with the two-parameter fitting would yield estimates $R = 0.717$, $u_{s,1}^{cr} = 0.225$. For the transition of loss of percolation of the healthy cells (Fig. 2), the same approach as illustrated in Fig. 6, yields estimates $R = 0.620$, $u_h^{cr} = 0.187$, $\beta = 0.424$. For the loss of percolation (the second transition) of the senescent cells (Fig. 2), we get $R = 0.651$, $u_{s2}^{cr} = 0.252$, $\beta = 0.416$. The fact that all our estimates of $\beta$ are close to the more precise uncorrelated-percolation literature value, indicates that, while our fitting approach has systematic errors, we can consider the results as a sufficient evidence that, as expected on general grounds, the universality class is the same.

Interestingly, for the onset of percolation transition of vacancies in the regime of interest described in connection with the loss of the long-duration steady state, Fig. 4, plots similar to Fig. 6 showed a much broader finite-size rounding of the transition, and as a result our simplified approach failed (there was no identifiable region for a convincing-quality least-squares power-law fitting). While there is no reason to expect a different universality class for vacant-site percolation transition(s) in the present model, a more careful study (with varying system sizes, periodic boundary conditions, and finite-size scaling analysis) is needed. However, this is outside the scope of the present work.

## 8. SUMMARIZING DISCUSSION

In summary, the main result of the present work is the conclusion that in 3D systems senescent cells can form well-connected clusters while healthy cells are being depleted and dying out. This property was observed with a minimal and very simple set of dynamical rules for cell populations evolution. The model can be significantly extended to study various processes and yield expressions for few-parameter fitting of data for measured tissue properties. These parameters are usually various process rates, which can thus be extracted and classified based on macroscopic measurements. Visualization of clusters, including those for heterogeneous, off-lattice and other model extensions can be used for comparison with experimental observations of tissue morphology.

After this article was submitted, a new important experimental work was published[58] reporting that mice with a transgene that ablates senescent cells, enjoy 35% longer lifespan and improved health. Specifically, the authors of Ref. 58 created an inducible transgene, which triggers activation of caspases and therefore apoptosis when bound by a drug,[2] AP20187. The authors of Ref. 58 further created a line of transgenic mice, which express above-mentioned transgene under $p16^{Ink4a}$ gene promoter, which is a universal marker of senescent cells. Treatment of these mice with AP20187, effectively eliminated naturally occurring senescent cells, which resulted in extended longevity and improved physiological parameters. These experimental findings are consistent with our computational results, suggesting a potential utility for our approach in modeling anti-aging therapies.

## ACKNOWLEDGEMENTS

We wish to thank Joan Adler, Sergii Domanskyi, and Charles M. Newman for useful input and collaboration. The work was in part supported by a seed grant from Cornell Center for Vertebrate Genomics 2014 to SL.

SUPPLEMENTARY INFORMATION

An illustrative movie showing spatial views of the just-formed infinite cluster (the one snapshotted in Fig. 2C).